\documentclass[aps,preprint,pra,superscriptaddress,amsmath]{revtex4-2}
\usepackage{graphicx,xcolor}

\begin{document}

\title{Influence of coherent vibronic excitation on the high harmonics generation of diatomic molecules}

\author{Mohammad Monfared}
\email[corresponding author: ]{m.monfared@yahoo.com}
\affiliation{Department of Physics, Faculty of Basic Sciences, Tarbiat Modares University, P.O. Box 14115-175, Tehran, Iran}
\affiliation{Inst. for Theoretical Physics,Vienna University of Technology, Wiedner Hauptstr. 8-10, A-1040 Vienna, Austria, EU}
\author{Elnaz Irani}
\email{e.irani@modares.ac.ir}
\affiliation{Department of Physics, Faculty of Basic Sciences, Tarbiat Modares University, P.O. Box 14115-175, Tehran, Iran}
\author{Christoph Lemell}
\affiliation{Inst. for Theoretical Physics,Vienna University of Technology, Wiedner Hauptstr. 8-10, A-1040 Vienna, Austria, EU}
\author{Joachim Burgd\"orfer}
\affiliation{Inst. for Theoretical Physics,Vienna University of Technology, Wiedner Hauptstr. 8-10, A-1040 Vienna, Austria, EU}

\date{\today}

\begin{abstract}
The generation of high harmonics (HHG) in atomic systems by the highly non-linear response to an intense laser field is a prominent pathway to the synthesis of ever shorter laser pulses at increasingly higher photon energies. Extensions of this process to molecules add to the complexity but also offers novel opportunities as multi-center effects and the coupling to nuclear degrees of freedom can influence HHG. In this work we theoretically explore the impact of coherent vibronic excitations of diatomic molecules on the HHG spectrum within the framework of time-dependent density functional theory. We observe the appearance of novel interference structures in the HHG spectra controlled by resonances between the driving field and the vibronic wavepacket.
\end{abstract}
\maketitle

\section{Introduction}\label{intro}
High harmonic generation (HHG) represents one of the major gateways towards obtaining spatially and temporally coherent extreme ultraviolet radiation light sources with unique properties for a wide range of applications \cite{corkum2007,krausz2009,ivanov2014,lepine2014,calegari2016,peng2019,li2020}. HHG has enabled the study of structural and dynamical information of matter and control of nuclear and electronic dynamics on their natural scale times \cite{solanpaa2014,ranitovic2014,lan2017}. The nonlinear process of HHG from atoms is well understood in terms of the three-step model \cite{corkum1993} in which electrons are first released from the ground state of the atom to the continuum by tunneling, then accelerated by the laser electric field, and finally recombine with the parent ion. Numerical simulations for this process have been performed by a large variety of theoretical methods for solving the time-dependent Schr\"odinger equation (TDSE) including the single-active electron (SAE) approximation \cite{a1,a11}, the grid-based exact numerical solution of the two-electron TDSE for helium \cite{a2,a3}, time-dependent density functional theory (TDDFT) \cite{Runge1984} applied to multi-electron atoms on the mean-field level \cite{a4}, the multi-configuration time-dependent Hartree Fock (MCTDHF) approach \cite{a5} including correlation effects beyond TDDFT, and, more recently, the time-dependent 2-particle reduced density matrix (TD-2RDM) method \cite{a6} bypassing the need for representing the $N$-electron wavefunction.

Extensions to molecular systems open new challenges as the coupling between electronic and nuclear degrees of freedom and multi-center interference effects have to be included. The extension of the three-step or Lewenstein model \cite{a7} to include nuclear degrees of freedom provides a helpful guide and detailed physical insight into the influence of the nuclear motion on the HHG process \citep{lein2005}. Early numerical simulations of HHG in molecular systems have employed the Born-Oppenheimer (BO) approximation, specifically in the fixed-ion approximation with the nuclei placed at the minima of the ground-state BO potential surface. Initially, the effect of nuclear motion on the HHG process has been theoretically studied only for small systems employing either the SAE approximation or reduced-dimension models for two-electron molecules (H$_2$, D$_2$) \cite{qu2001,lein2005,madsen2006,bandrauk2008,bandrauk2009,liu2010,ahmadi2014}. These theoretical studies predict that the nuclear motion leads to a reduction of harmonics yield \cite{lein2005,madsen2006}. The signature of nuclear motion has been seen in the time profiles of high-order harmonics from H$_2$ by Bandrauk et al.\ \cite{bandrauk2008}. For high laser intensities ($I\sim 1$ PW/cm$^{2}$) the nuclear motion shortens that part of the attosecond pulse train originating from ionization of the first electron and facilitates the onset of the contribution from the release of the second electron for longer pulses. For lower laser intensities ($I\sim 0.1$ PW/cm$^{2}$), electron excitation due to recollision of the returning electron was observed in the time profile of the attosecond pulse \cite{bandrauk2009}. Comparing the experimental HHG signals from H$_2$ with those from heavier molecules (N$_2$) as well as the atomic HHG signal from Ar all three systems of which feature very similar ionization potentials has demonstrated that, indeed, the nuclear motion between the ionization and recombination steps leads to an effective suppression of the HHG yield and, in addition, to a broadening of the harmonic signal \cite{yuan2016}. Moreover, retrieving the nuclear motion of H$_2$ and D$_2$ from the harmonic spectra with sub-fs time resolution has been demonstrated theoretically \cite{lein2005} and experimentally \cite{baker2006,lan2017}. Nuclear motion reconstruction from experimental HHG signals can be done either by the ratio of harmonics intensity \cite{baker2006} or the observed frequency shift in HHG signals induced by nuclear motion \cite{lan2017}. TDDFT allows the simulation of HHG for large multi-electron molecules \cite{Chu2001,baer2003,Telnov80,Petrelli2010, Bandrauk2011,Dundas2012}. While most earlier applications used the fixed-ion approximation, more recent implementations have accounted for nuclear motion by solving concurrently the classical equations of motion for the ions within the framework of Ehrenfest dynamics \cite{Kunert2003,Dundas,Calvaynac2000, Castro2004,wardlow2016,mulholland2018}.

In a pioneering experimental study, Li et al.\ \cite{li2008} explored in a pump-probe setting the role of molecular motion in HHG in the presence of vibronic excitations rather than in the ground state. Exciting the N--N stretch mode of the N$_2$O$_4$ molecule they could identify the dependence of the HHG on the phase of vibration with a peak of HHG emission occurring near the outer turning point of the vibration. In the present work, we theoretically study the effects of the excitation of vibronic wavepackets on the HHG spectrum of several diatomic molecules (H$_2$, N$_2$, F$_2$, HF). We treat the problem in its full dimensionality and include multi-electron effects on the level of TDDFT. The quasi-classical motion of the vibronic wavepacket is described by Ehrenfest dynamics. Quantum effects are included by coherently averaging the harmonic signal over the wavepacket. One of our key findings is that novel resonance structures appear in the HHG spectra of molecules when the frequencies of vibronic motion and of the infrared (IR) laser field driving the electronic dynamics are commensurate to each other.

In section \ref{meth} we briefly review the methods employed in our simulations. A comparison between HHG spectra with and without accounting for the nuclear degrees of freedom are given in Sec.\ \ref{results}. The case of resonant driving by commensurate frequencies between laser field and coherently excited vibronic wavepackets is investigated in Sec.\ \ref{results2}. Conclusions drawn from our results are discussed in Sec.\ \ref{end}. Atomic units ($e=m_e=\hbar=1$ a.u.) are used throughout the paper unless stated otherwise.

\section{Simulation Methods}\label{meth}
We model the interaction of strong laser pulses with molecules within the framework of time dependent density functional theory employing the software package \textsc{OCTOPUS} \cite{octopus1,octopus2}. This also allows for the concurrent solution of Newton's equations of motion for the ionic cores in the combined external laser field and the electrostatic potential of the time dependent electronic density $n(\vec r,t)$ thereby including the coupling between electronic and nuclear degrees of freedom on a classical level. For a spin-compensated system containing $N$ electrons the electronic density is given by
\begin{equation}
n(\vec r,t)=2\sum_{i=1}^{N/2} |\psi_i(\vec r,t)|^2,
\end{equation}
where the $\psi_i(\vec r,t)$ are the time dependent Kohn-Sham (TDKS) orbitals, the solutions of the TDKS equations
\begin{equation}\label{TDKS}
i\frac{\partial}{\partial t}\psi_i(\vec r,t)=\left[ -\frac{1}{2}\nabla^2 + V_\mathrm{H}(\vec r,t) + V_\mathrm{xc}(\vec r,t) + V_\mathrm{ext}(\vec r,\vec R_1,\vec R_2,t) \right] \psi_i(\vec r,t)\, .
\end{equation}
In Eq.\ \ref{TDKS}, $V_\mathrm{H}(\vec r,t)$ and $V_\mathrm{xc}(\vec r,t)$ are the Hartree and exchange-correlation potentials, respectively. $V_\mathrm{ext}(\vec r,\vec R_1,\vec R_2,t)$ is the external potential representing the Coulomb attraction between electrons and nuclei located at positions $\vec R_\alpha$ and the interaction potential with the laser pulse. In dipole approximation and length gauge $V_\mathrm{ext}$ is given by
\begin{equation}
V_\mathrm{ext}(\vec r,\vec R_1,\vec R_2,t)=V_{ne}(\vec r,\vec R_1,\vec R_2,t)+\vec E(t)\cdot\vec r\, .
\end{equation}
As $V_{ne}$ we use for H$_2$, N$_2$, and HF Hartwigsen-Goedecker-Hutter LDA pseudopotentials \cite{hartwigsen1998}, for F$_2$ an optimized Norm-Conserving Vanderbilt PBE pseudopotential \cite{schlipf2015}. The linearly polarized laser field is aligned parallel to the molecular axes and is described in terms of the vector potential by
\begin{equation}
\vec A(t)=A_0\, f_\mathrm{env}(t) \sin[\omega_L(t-t_0)]\, \hat e_x
\end{equation}
with $A_0$ the peak field, $f_\mathrm{env}(t)$ the pulse-envelope function, and $\omega_L$ the laser frequency. The electric field follows from
\begin{equation}
\vec E(t)=-\frac{1}{c}\frac{\partial\vec A}{\partial t}\, .
\end{equation}
We use pulses with a trapezoidal envelope with one cycle ramp-on, 4 to 12 cycles constant amplitude, and one cycle ramp-off. The typical peak intensity is $I_0=10^{14}$ W/cm$^2$ (peak amplitude $E_0=0.053$ a.u.), and we probe the dependence on the infrared wavelength $\lambda$. The polarization unit vector $\hat e_x$ is aligned with the molecular axis.

Two different functionals for $V_\mathrm{xc}[n(\vec r,t)]$ are used in the present work. While for most molecules studied in this work (H$_2$, HF, N$_2$) the self-interaction corrected local density approximation (SIC-LDA) to the  exchange-correlation potential in the parametrization by Perdew and Zunger \cite{perdew81} could be used to satisfactorily reproduce both electronic properties \textit{and} the interatomic potential (equilibrium distance, vibrational frequency) a more complex pathway had to be taken to model HHG for F$_2$ molecules. Simultaneously reproducing electronic and vibronic properties using the same functional has been found to be difficult for this molecule. We therefore proceed in two steps: In the first step, we calculated the vibronic properties of the fluoride atoms using the generalized gradient approximation (GGA) \cite{perdew96} to $V_\mathrm{xc}$ without self-interaction correction which yields reasonable agreement with experimental data. In the second step, the electron dynamics were modeled using the SIC-GGA functional in the combined external laser field and the atomic Coulomb field derived from the atomic positions recorded in the first step as the SIC-GGA functional performs considerably better for the electronic properties. We refer to this combined approach as GGA/SIC-GGA. Since high harmonic generation probes predominantly the short-time response of the molecule well within one vibrational period, the application of this GGA/SIC-GGA approximation appears to be justified.

In order to prevent unphysical reflections of the liberated electrons from the boundary of the computing box, the KS-orbitals $\psi_i(\vec r,t)$ are multiplied with a masking function which is unity in the inner simulation region and is gradually switched to zero at the borders. The width of the masking function is set to 40 a.u.\ in the direction of the laser polarization and 10 a.u.\ for the perpendicular directions. Convergence of the calculations with respect to grid parameters and time step have been thoroughly tested and confirmed.

The ionic motion under the influence of the ion-electron potential, the internuclear Coulomb repulsion, and the interaction of the nuclei with the external laser field is treated classically (``Ehrenfest TDDFT'')
\begin{equation}
M_{1,2}\ddot{\vec R}_{1,2}=-\int n(\vec r,t)\frac{V_{ne}(\vec r,\vec R_1,\vec R_2,t)}{\partial R_{1,2}}\, d\vec r - \frac{\partial V_{nn}(\vec R_1, \vec R_2)}{\partial R_{1,2}} + Z_{1,2}\vec E(t) \label{nucl-mot}
\end{equation}
with $V_{nn}$ the (shielded) Coulomb repulsion between the nuclei. In Eq.\ \ref{nucl-mot} $M_\alpha$ and $Z_\alpha$ are the masses and core charges of the nuclei, respectively. Neglecting the slow rotational degrees of freedom, the HHG spectrum is derived from the induced time dependent dipole acceleration $a=\ddot{d}$. The dipole moment
\begin{equation}
d[R(t),t]=\int d^3r\, n[\vec r, R(t), t]\, x\, ,\label{dipole}
\end{equation}
is a function of the time-dependent internuclear distance $R(t)=|\vec R_1-\vec R_2|$ of the diatomic molecule. As point of reference for the following calculations which include the vibrational degrees of freedom we also perform calculations for fixed internuclear distances $R(t)=R$ within which the induced dipole depends only parametrically on $R$. In most cases we choose for the latter the equilibrium distance $R=R_\mathrm{eq}$ of the ground-state Born-Oppenheimer potential curve.

Going beyond the standard Ehrenfest-TDDFT to account for quantum effects we include the shape and spatial width of the vibrational wavepacket by extending the approach of Chu and Groenenboom \cite{chu2012} originally introduced for vibronic ground states to the present case of coherent wavepackets. Accordingly, we follow the Ehrenfest dynamics for the center of the vibronic wavepacket $\chi(R,t)$ localized near
\begin{equation}
R_0(t)=R_\mathrm{eq}+\delta R(t)
\end{equation}
with the time-dependent variation $\delta R(t)$ calculated from Eq.\ \ref{nucl-mot}. In order to sample the vibrational wavefunction we solve in addition the TDDFT equation for $n(t)$ for an ensemble of nearby classical but non-Ehrenfest trajectories $i$,
\begin{equation}
R_i(t)=R_0(t)+\Delta R_i\, ,
\end{equation}
with $\Delta R_i=i\sigma$ and $i=-M, -M+1, \ldots, M$ to cover the spatial extent of the time-dependent vibronic wavepacket with width $\sigma$ (in the present case we take $M=2$). Inserting the resulting $R_i(t)$ dependent electronic densities into the dipole moment (Eq.\ \ref{dipole}) yields the Fourier transform of the dipole acceleration
\begin{equation}
a[R_i(0),\omega] = -\omega^2d[R_i(0),\omega]\label{eq10}
\end{equation}
with
\begin{equation}
d[R_i(0),\omega]=\frac{1}{T}\int_0^T dt\, e^{-i\omega t}d[R_i(t),t]
\end{equation}
which is parametrically dependent on the initial value $R_i(0)$. The coherent average over the ensemble generated by the initial conditions $\{R_i(0),i=-M,...,M\}$ represents now the evolved vibrational wavefunction $\chi(R)$ and is accordingly given by
\begin{equation}
\langle a(\omega)\rangle = \sum_{i=-M}^M \Delta R |\chi(R_i)|^2 a[R_i(0),\omega]\, .\label{eq11}
\end{equation}
The HHG spectrum follows from
\begin{equation}
S(\omega)=\frac{3}{2\pi c^3}|\langle a(\omega)\rangle|^2\, .\label{eq12}
\end{equation}
In the limit $\delta R(t)=0$, i.e., for fixed ions, Eqs.\ \ref{eq11} and \ref{eq12} reduce to the result (Eq.\ 14) of Ref.\ \cite{chu2012} for the vibrational ground state. For a non-zero $\delta R(t)$, Eqs.\ \ref{eq11} and \ref{eq12} allow for the inclusion of quantum effects into the vibrational contribution to HHG while simultaneously employing Ehrenfest TDDFT. Further insight into the timing of the generation of harmonics and its dependence on the nuclear motion can be gained by a time-frequency analysis computing the Gabor transform (windowed Fourier transform) of the dipole acceleration,
\begin{equation}
\vec A(t,\omega)=\int_{-\infty}^\infty \ddot{\vec d}(t')\exp(-i\omega t')\exp\left[\frac{(t'-t)^2}{2\sigma_\tau^2}\right] dt'\, .
\end{equation}

\begin{figure}[h!]
\centerline{\includegraphics[width=8cm]{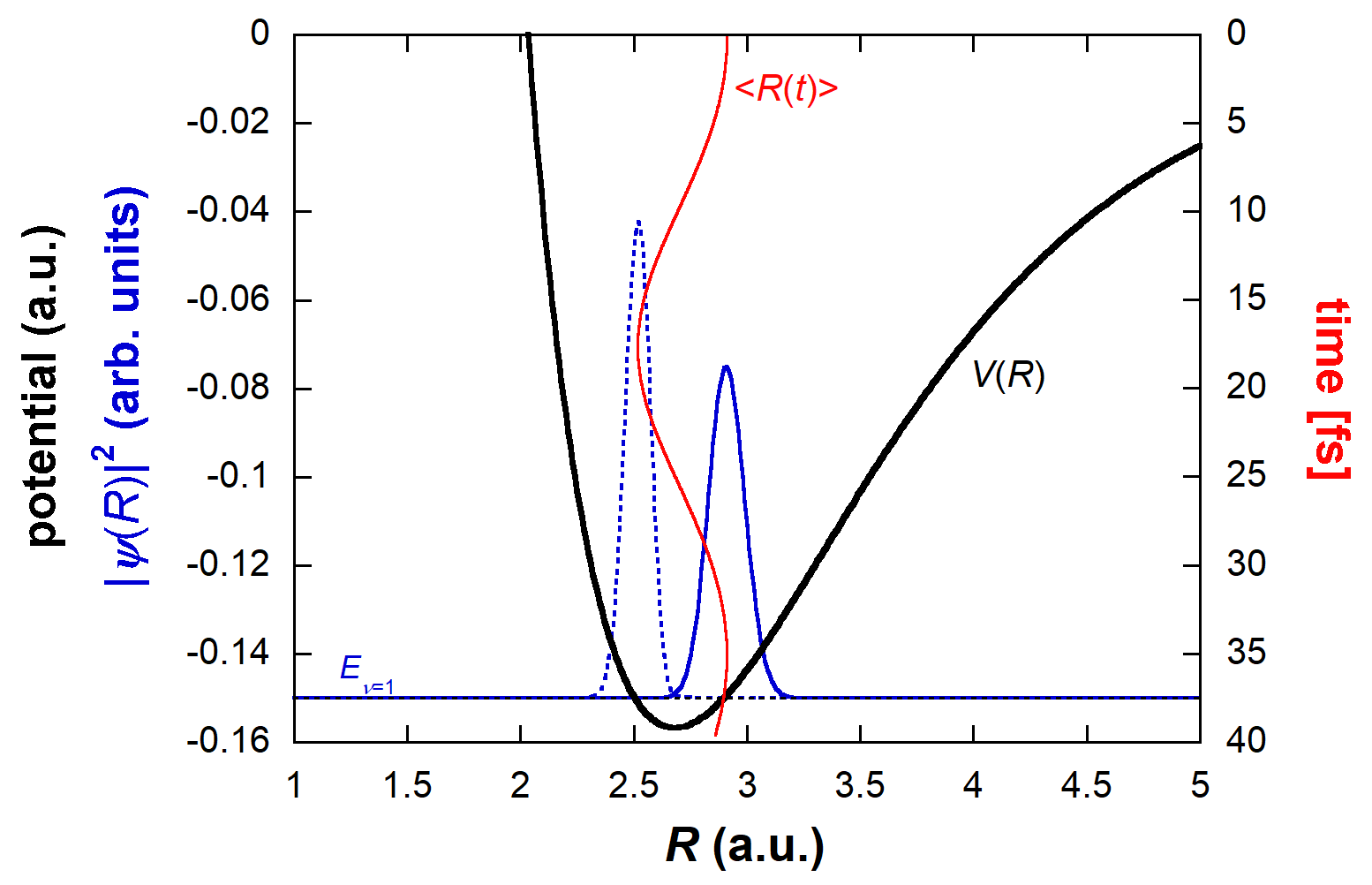}}
\caption{Morse-type potential of F$_2$ (thick black line) and coherent vibronic wavepacket at turning points (dashed and solid blue lines). Temporal evolution of expectation value $\langle R(t)\rangle$ (red solid line; time in fs on right axis).\label{fig1}}
\end{figure}

In the following we will apply Eqs.\ \ref{eq11} and \ref{eq12} to the present case of coherently excited vibronic wavepackets. Such wavepackets can be excited by a wide variety of techniques including impulsive collisional excitation, Franck-Condon transitions, or by impulsive stimulated Raman scattering \cite{li2008}. Close to the potential minimum of the BO curve the vibronic wavepacket can be approximated by a Schr\"odinger-like dispersion-free coherent state \cite{schrodinger1926} of the harmonic oscillator subsequently extensively employed in the field of quantum optics \cite{glauber1963}
\begin{equation}
|\alpha(t)\rangle = e^{-i\Omega t/2}e^{-|\alpha|^2/2}\sum_{\nu=0}^\infty \frac{\alpha(t)^n}{\sqrt{n!}}|\chi_\nu\rangle\label{eq13}
\end{equation}
with $\alpha(t)=|\alpha|e^{-i(\Omega t+\phi)}$ and $|\alpha |=|\delta R(t)|_\mathrm{max}/\sigma$ the scaled amplitude of excitation. $|\alpha|$ can be used to parametrize the amplitude of coherent excitation, $|\alpha|=\sqrt{\langle E\rangle/\Omega} \geq \frac{1}{\sqrt{2}}$. Generalization of Eq.\ \ref{eq13} to anharmonic Morse potentials approximating the BO potential surface for higher excitations is straight forward \cite{Morse1,Morse2,Kais1990,Angelova2008}. An example of a (near) minimum uncertainty vibronic wavepacket is shown in Fig.\ \ref{fig1}. To first order, anharmonic corrections can be accounted for by a time-dependent width $\sigma(t)=\sigma[1+c\,\sin(\Omega t+\phi)]$ of the quantum wavepacket the oscillation of which being synchronized with the vibration. We have performed wavepacket-propagation simulations in the internuclear potential to extract the values for $\sigma(t), c$, and $\phi$ for the molecules studied here. For such quasi-classical coherent vibronic wavepackets, Ehrenfest-TDDFT appears to be well-suited to account for the interplay between electronic and the quasi-classical ionic dynamics.

\section{Comparison Between Fixed and Moving Nuclei}\label{results}
In this section we present the results of our simulations performed for the homonuclear molecules H$_2$, N$_2$, and F$_2$ and the heteronuclear molecule HF. We start with examining the self-consistently determined groundstate parameters and discuss the generation of high harmonics in molecules with fixed and moving nuclei. While TDDFT is appropriate for multi-electron molecules, its applications to H$_2$ primarily serve to illustrate the interplay between the electronic and ionic dynamics for the fast and large-amplitude oscillations of protons.

\subsection{Equilibrium conditions}\label{eq-cond}
We have first tested the accuracy of the static DFT predictions for ground-state properties of the investigated molecules using the same functionals as in the subsequent TDDFT simulations. A geometry-optimization calculation was performed to obtain the relaxed coordinates for the cores which are then used to determine the electronic groundstate as well as the internuclear (Morse-like) potential. As can be seen from table \ref{tab:table1} experimental equilibrium distances and vibration frequencies are in most cases reasonably well reproduced using the SIC-LDA exchange-correlation potential. Equilibrium distances $R_\mathrm{eq}$ are found to be slightly smaller and, correspondingly, vibrational frequencies $\Omega$ slightly larger in agreement with previous calculations \cite{wardlow2016}. However, large deviations are found for F$_2$. For this molecule we performed a combined two-step GGA/SIC-GGA calculation. We first determined the molecular parameters using the GGA functional neglecting SIC and subsequently performed the electronic structure calculation for these molecular structure parameters using the SIC-GGA functional. The observed minor differences ($\lesssim 4\%$) in the binding energies of the highest occupied molecular orbitals (HOMO) has some influence on the absolute harmonic yield but does not change the structure of the HHG spectra discussed below.

\begin{table}[h!]
\caption{\label{tab:table1}%
Comparison between calculated and measured values for equilibrium distances, vibrational frequency, and binding energies of the highest occupied molecular orbitals (HOMO) for the diatomic molecules used in this study. All values are given in a.u.
}
\begin{ruledtabular}
\begin{tabular}{rccccc}
&$R_\mathrm{eq}$&$\Omega$&HOMO&HOMO-1&HOMO-2\\
\colrule
\multicolumn{6}{c}{H$_2$}\\
Exp.&1.400&2.01e--2&0.5669& & \\
SIC-LDA&1.311&2.26e--2&0.6834& & \\
\multicolumn{6}{c}{N$_2$}\\
Exp.&2.075&1.06e--2&0.5726&0.6233&0.6883\\
SIC-LDA&1.948&1.33e--2&0.6058&0.6918&0.6992\\
\multicolumn{6}{c}{F$_2$}\\
Exp.&2.668&4.18e--3&0.5832&0.6910&0.7750\\
SIC-LDA&2.152&7.2e--3&0.5428&0.8275&0.9682\\
GGA&2.68&4.38e--3&0.3471&0.4697&0.5696\\
GGA/SIC-GGA&&&0.5681&0.7034&0.8139\\
\multicolumn{6}{c}{HF}\\
Exp.&1.732&1.88e--2&0.5898&0.7284&1.4545\\
SIC-LDA&1.667&1.99e--2&0.6212&0.7613&1.3665
\end{tabular}
\end{ruledtabular}
\end{table}

\begin{figure}[h!]
\centerline{\includegraphics[width=16cm]{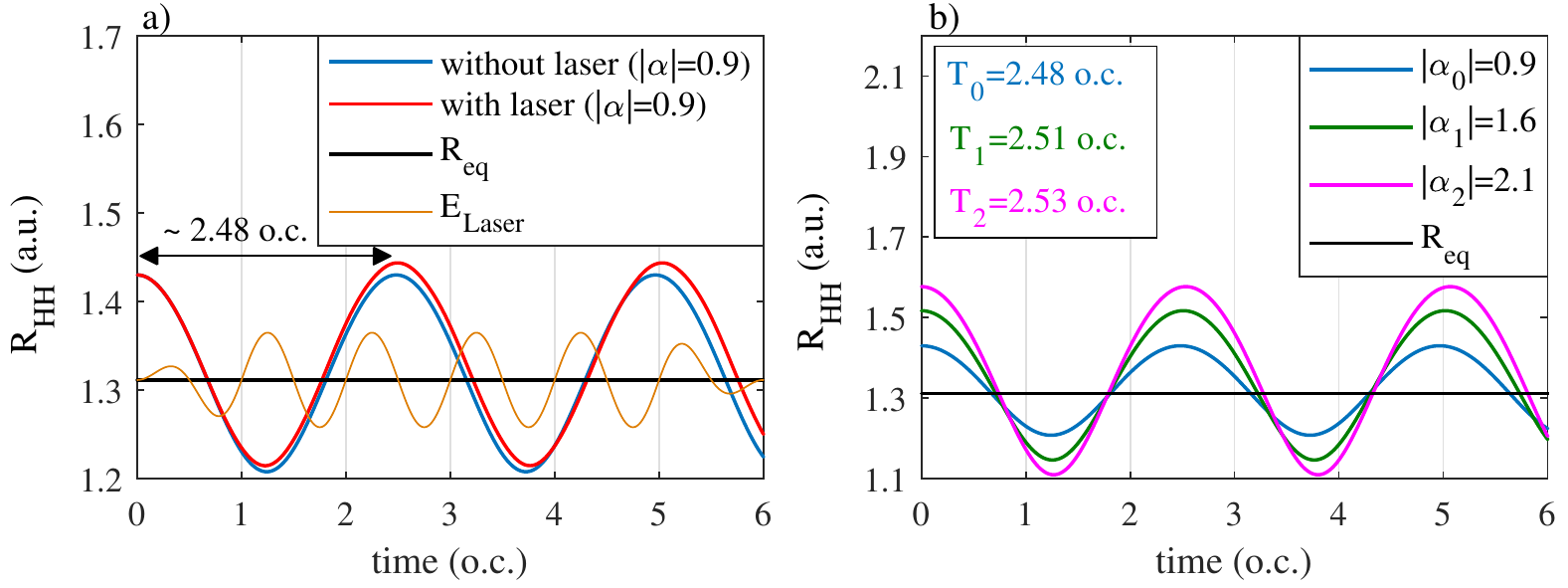}}
\caption{(a) Comparison between the time-dependent internuclear distance $R(t)$ of H$_2$ in the presence (absence) of the IR laser field ($\lambda = 800$ nm, $I_0=10^{14}$ W/cm$^2$) as a function of time in units of optical cycles (o.c.). The amplitude of the laser field is displayed for illustrative purposes in arbitrary units. For later reference we note that the frequencies are approximately commensurate, $\omega_L/\Omega=5/2$. The equilibrium distance $R_\mathrm{eq}=1.31$ a.u.\ predicted by SIC-LDA is shown for reference. (b) Time dependent $R(t)$ for field-free evolution predicted by Ehrenfest-TDDFT for increasing excitation amplitude $\delta R(t=0)=|\alpha_k|\cdot\sigma$ with periods $T_k$.\label{fig2}}
\end{figure}

The prediction of Ehrenfest-TDDFT for the internuclear distance $R(t)$ for H$_2$ as a function of time in the presence of a driving laser field is compared with the field-free motion in Fig.\ \ref{fig2}. For this light molecule the interaction with the laser slightly increases the coherent vibrational amplitude and, correspondingly, decreases the oscillation frequency (Fig.\ \ref{fig2}a). The latter is an immediate consequence of the anharmonicity explored for increased initial excitation amplitudes $\delta R(t=0)=|\alpha|\cdot\sigma$ in the absence of a driving field (Fig.\ \ref{fig2}b). The increasing anharmonicity is accompanied by an increase of the mean $\bar R$, the temporal average over $R$ over one oscillation period. For heavier nuclei (e.g.\ in F$_2$) the laser induced change of the vibronic frequency becomes negligible for an ultrashort few-cycle pulse. This observation also \textit{a posteriori} justifies the use of different exchange-correlation potentials for the nuclear and electronic degrees of freedom for F$_2$ even so it breaks the self-consistency of the coupling between nuclear and electronic motion. For all other molecules the non-linear electronic response has been fully self-consistently calculated including the influence of the external laser field and the time-dependent electronic density fluctuations on the nuclear motion.

\subsection{High Harmonic Spectra}
\begin{figure}[h!]
\centerline{\includegraphics[width=16cm]{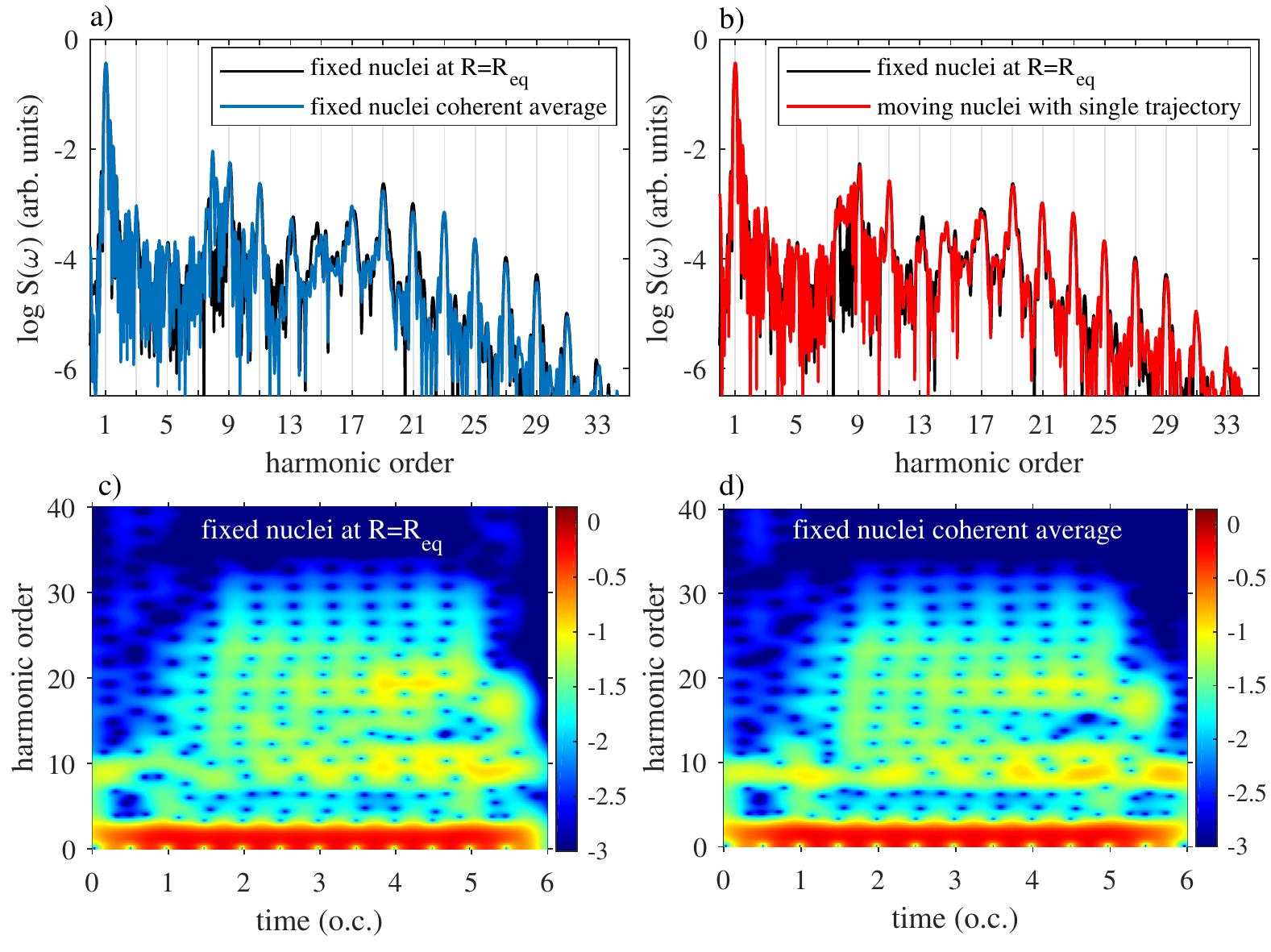}}
\caption{Harmonic spectrum of F$_2$ driven by a 6-cycle laser pulse ($\lambda=800$ nm, $I_0=10^{14}$ W/cm$^2$). (a) Comparison between fixed nuclei at $R=R_\mathrm{eq}$ and sampling over the stationary ground-state wavefunction (Eq.\ \ref{eq11}) with $\delta R(t)=0$. (b) Comparison between $R=R_\mathrm{eq}$ and a single Ehrenfest trajectory starting at the outer turning point $R_0(t=0)=R_\mathrm{eq}+|\alpha|\sigma$ with scaled amplitude $|\alpha| = 0.7$. (c) Time-frequency analysis of HHG from F$_2$ for fixed nuclei (blackline in panel a) and (d) for the coherent average in the fixed nuclei approximation (blue line in panel a). The temporal width of the Gaussian window function is $\sigma_\tau=0.5$ fs ($\approx 0.19$ o.c.).\label{fig3}}
\end{figure}

A comparison between the resulting high-harmonic spectra of F$_2$ when the nuclear degree of freedom is taken into account at  various levels of approximation is presented in Fig.\ \ref{fig3}. The calculation at fixed nuclear distance at the equilibrium position $R=R_\mathrm{eq}$ agrees quite well with the one when the coherent average over the vibrational ground-state wavefunction is taken into account (Eq.\ \ref{eq11} with the limit $\delta R(t) = 0$; Fig.\ \ref{fig3}a). This indicates that, overall, the modulus and phase of the dipole acceleration amplitude $a(R,\omega)$ (Eq.\ \ref{eq10} calculated for fixed $R$ varies relatively little with $R$ over the extent of the vibrational wavefunction. A pronounced exception appears near harmonic $8\omega$. Here we observe an enhancement of the spectrum. This frequency nearly matches the HOMO-LUMO gap in F$_2$ which becomes accessible as $R$ is parametrically varied leading to an enhancement of $a(R,\omega)$ which is absent in the calculation at fixed $R=R_\mathrm{eq}$. The width of this peak is directly related to the time dependent change of the HOMO-LUMO gap energy during the vibrational motion but, unlike the harmonic peaks, features no sub-(optical) cycle structure (Figs.\ \ref{fig3}c,d). The latter is a hallmark of electron emission, acceleration, and recombination during $\sim 3/4$ of an optical cycle. The spectrum also closely resembles the spectrum for a single Ehrenfest-TDDFT trajectory with a time-varying $R(t)$ (Eq.\ \ref{nucl-mot}) with amplitude $|\alpha|\approx 0.7$ initialized at the outer turning point (Fig.\ \ref{fig3}b). 

\begin{figure}[h!]
\centerline{\includegraphics[width=16cm]{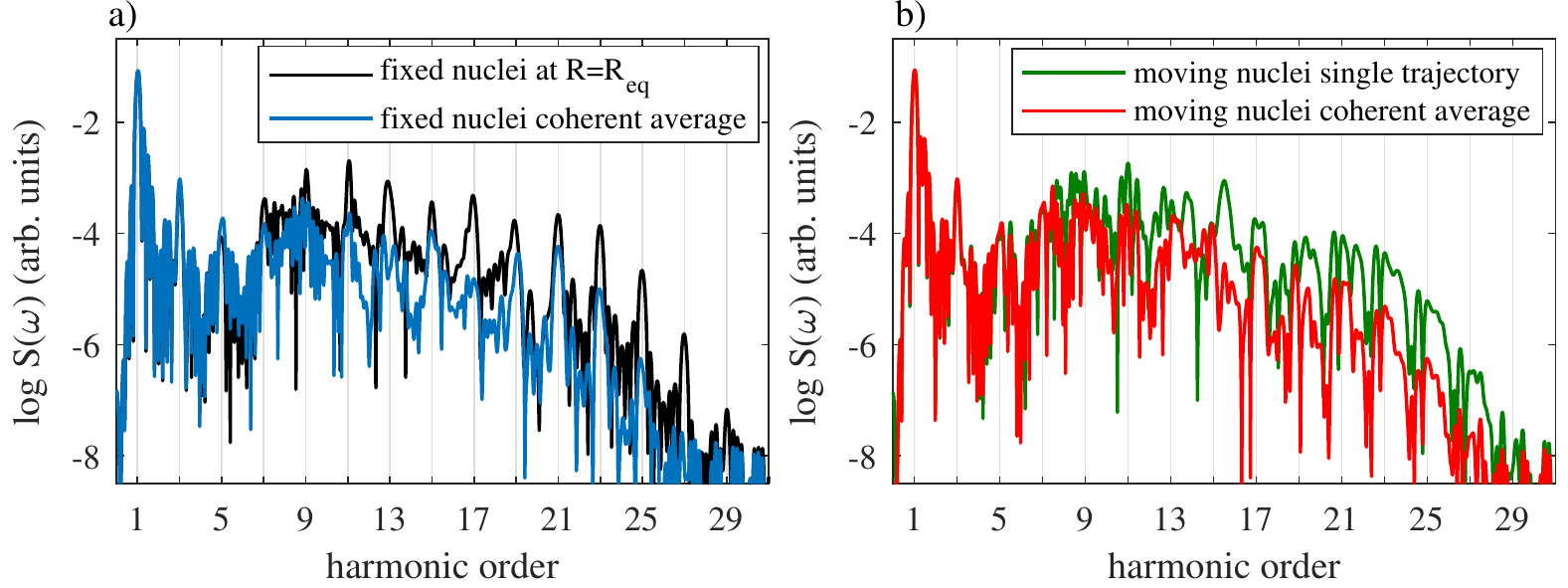}}
\caption{Harmonic spectra of H$_2$ driven by a 6-cycle laser pulse ($\lambda=742$ nm, $I_0=10^{14}$ W/cm$^2$). (a) Comparison between spectra for fixed nuclei $R=R_\mathrm{eq}$ and including sampling of stationary ground-state wavefunction. (b) Comparison between single Ehrenfest trajectory starting at the outer turning point $R_0(t=0)=R_\mathrm{eq}+|\alpha|\sigma$ with $|\alpha| = 0.9$ and sampling over a Schr\"odinger minimum uncertainty wavepacket.\label{fig4}}
\end{figure}

For the lightest molecule, H$_2$, the influence of the nuclear motion on the electronic dynamics is expected to be most pronounced. Indeed, sampling the stationary vibrational ground-state wavefunction (Eq.\ \ref{eq11} with $\delta R(t)=0$) reduces the harmonic spectrum considerably compared to the fixed ion result with $R=R_\mathrm{eq}$ (Fig.\ \ref{fig4}a). The reduction becomes more pronounced with increasing harmonic order in agreement with the results of Ref.\ \cite{chu2012}. Comparing the spectrum for a single Ehrenfest trajectory $R_0(t)$ with $R_0(t=0)=R_\mathrm{eq}+|\alpha|\sigma$ with the spectrum including the sampling over a Schr\"odinger minimum uncertainty vibronic wavepacket (Eq.\ \ref{eq11} with $\delta R(t)\neq 0$; Fig.\ \ref{fig4}b) shows that the reduction of the HHG spectrum with increasing harmonic order persists. Overall, the effect of inclusion of quantum effects for the ground state and for coherent vibrational excitation spectra agree quite well lending credence to the present level of inclusion of vibrational dynamics.

\section{Spectra for commensurate frequencies}\label{results2}
When the frequency of the coherent vibronic wavepacket $\Omega$ is commensurate with the frequency $\omega_L$ of the laser driving the electronic dynamics, the coupling between the ionic and electronic dynamics is expected to be enhanced. This does not only apply to the case of a $1:1$ resonance, i.e.\ $\Omega=\omega_L$, which would require mid infrared driving fields but also higher-order resonances of commensurate frequencies, $\omega_L/\Omega=m/n$ ($m,n$: integers). For H$_2$ and $\lambda=800$ nm, this frequency ratio happens to be close to $5:2$. For N$_2$ and $\lambda=1147$ nm, a $3:1$ resonance can be reached. We explore in the following the impact of driving with commensurate frequencies on the high-harmonic spectrum and irradiate N$_2$, H$_2$, and HF with laser pulses with commensurate frequencies. The peak intensity was set to $10^{14}$ W/cm$^2$, the wavelengths of the resonant laser pulses were 800 nm for a $5:2$ resonance in H$_2$, 1147 nm for a $3:1$ resonance in N$_2$, and 800 nm for a $3:1$ resonance in HF, a polar molecule for which also an optical excitation of a vibronic wavepacket appears feasible.

\begin{figure}[h!]
\centerline{\includegraphics[width=16cm]{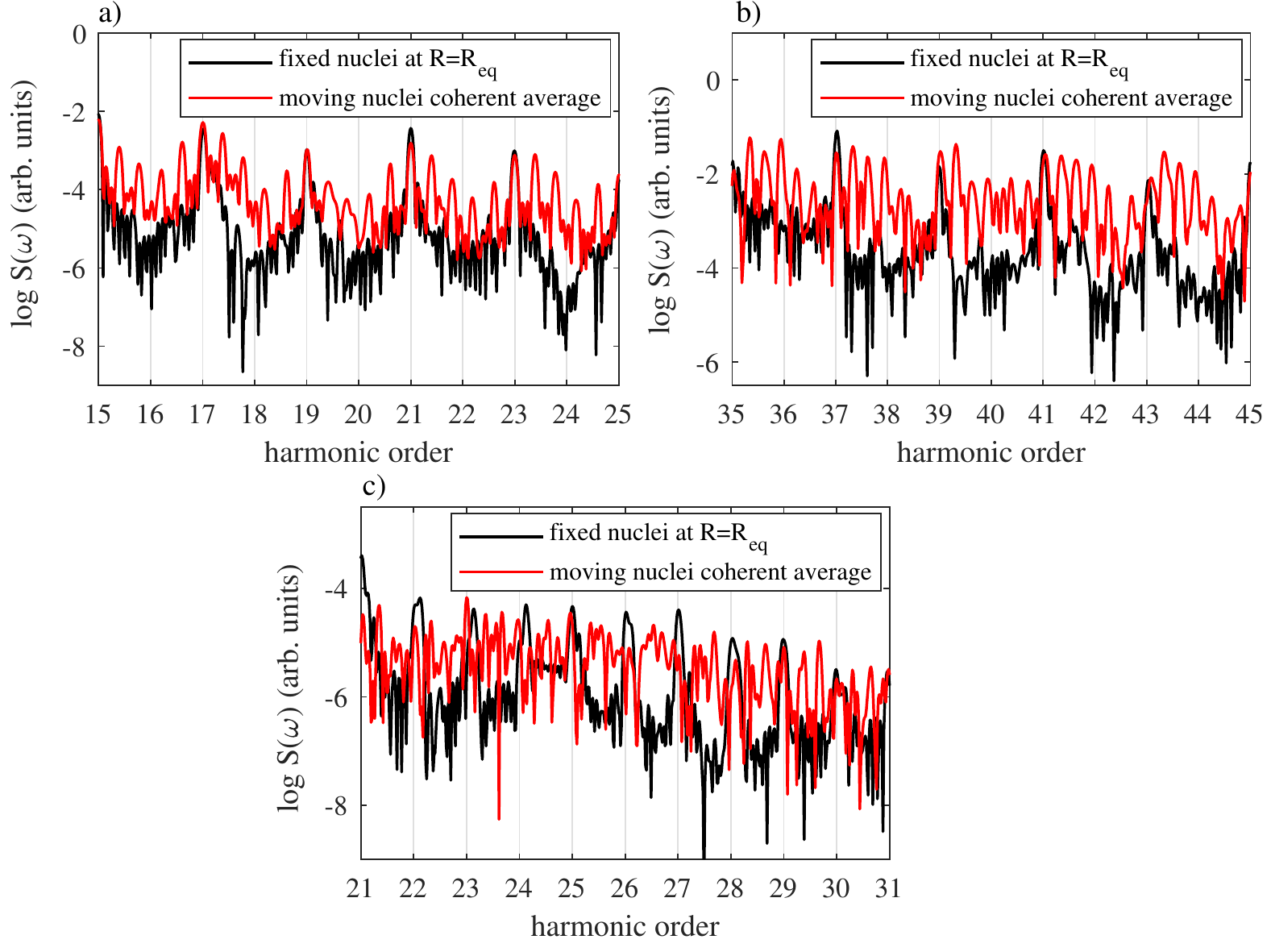}}
\caption{Combined vibronic-electronic harmonics for driving frequencies commensurate with the vibronic wavepacket. (a) H$_2$, (b) N$_2$, and (c) HF. In each frame we compare the HHG spectrum for fixed internuclear distance $R=R_\mathrm{eq}$ with the result for the vibronic wavepacket treated by Ehrenfest-TDDFT including sampling over the width of the vibronic wavefunction (Eq.\ \ref{eq11} with $\delta R(t)\neq 0$). (a) H$_2$ with $\lambda=800$ nm, $I_0=10^{14}$ W/cm$^2$ and $|\alpha| = 1.8$, (b) N$_2$ with $\lambda=1147$ nm, $I_0=10^{14}$ W/cm$^2$ and $|\alpha| = 7.8$, (c) HF  with $\lambda=800$ nm, $I_0=10^{14}$ W/cm$^2$ and $|\alpha| = 3.25$. \label{fig5}}
\end{figure}

For this simulation we have increased the pulse length from 6 cycles to 14 cycles in order to narrow the spectral distribution of the pulse. For such longer laser-pulse interaction also the spectral width of the interference fringes become narrower and can be clearly identified up to the cut-off region of the spectrum (Fig.\ \ref{fig5}). The harmonic spectrum is fundamentally altered. As the emission of harmonics from recollision events distributed over several optical cycles adds up coherently the information on the motion of the vibronic wavepacket directly enters the spectrum. The energy spacing of $2\hbar\omega_L$ between successive HHG peaks is now filled by additional peaks spaced by $\hbar\Omega$. Fig.\ \ref{fig5}a shows the appearance of the additional peaks for a single Ehrenfest trajectory with initial excitation $R_0(t=0)=R_\mathrm{eq}+|\alpha|\sigma$ which are completely absent in the fixed-ion approximation with $R=R_\mathrm{eq}$. It is important to note that the anharmonicity and resulting deviation from equispaced vibronic levels with spacing $\hbar\Omega$ does not destroy the resonance effect as long as the spectral mismatch due to the anharmonicity lies within the linewidth of the driving pulse determined by its duration. Inclusion of the sampling over a minimum-uncertainty vibronic wavepacket reduces the height of these combined vibronic-electronic harmonics but leaves the structure of the spectrum unchanged. The experimental observation of such a commensurate harmonic spectrum appears feasible. Even when the subpeaks can not be individually resolved, this resonance effect may facilitate an overall increase in the HHG yield. For N$_2$ and a higher excitation amplitude ($|\alpha| \approx 8$) similar vibronic-electronic harmonics for a frequency ratio $3:1$ can be found (Fig.\ \ref{fig5}b). For heavy nuclei, quantum effects due to the coherent averaging over the wavefunction of the vibrational wavepacket lead, unlike for H$_2$, only to a small change in the spectrum of F$_2$ (not shown). Pronounced vibronic-electronic harmonics can be also observed for the strongly polar molecule HF (Fig.\ \ref{fig5}c). Here we observe for $\lambda=800$ nm a $3:1$ resonance. Realization of such a resonance in this molecule would be facilitated by the property that the vibrational wavepacket could be optically excited and steered.

\section{Conclusions}\label{end}
We have investigated the influence of vibronic motion on the high-harmonics generation in several small molecules. We employed an extended Ehrenfest-TDDFT within which the ionic motion follows the classical equations of motion with forces self-consistently determined from the time-dependent electronic density propagated by TDDFT, quantum effects of the vibronic degree of freedom are included. The latter is accomplished by coherently sampling the harmonic amplitude over the time-dependent wavefunction of a minimum-uncertainty wavepacket. As expected, the influence of the vibronic degree of freedom is, in general, reduced with increasing mass of the molecular constituents. However, prominent resonance effects have been identified for commensurate frequencies whose appearance are independent of the atomic masses involved. When a multiple of the frequency $\omega_L$ of the laser driving the strong-field ionization and recombination of the electron in the molecule is in resonance with a multiple of the vibronic frequency, i.e.\ $\omega_L/\Omega=m/n$, the standard harmonic spectra with spacing $2\hbar\omega_L$ between adjacent peaks is drastically altered and replaced by a densely spaced spectrum with additional peaks spaced by $\sim\hbar\Omega\approx \frac{n}{m}\omega_L$. For these resonances we observe, overall, an increase of the integrated HHG intensity. A moderate anharmonicity of the vibronic spectrum does not destroy these resonance effects as long as the frequency mismatch lies within the Fourier width of the few-cycle infrared driving pulse. Experimental realization employing, for example, the impulse excitation of vibronic wavepackets \cite{li2008} appears within reach.

\acknowledgments This work has been supported by the doctoral college FWF-1243 (Solids4Fun) of the Austrian Science Fund FWF. The authors acknowledge the financial sponsorship provided by cofounding of Kharazmi University and Ministry of Science, Research, and Technology (MSRT) of the Islamic Republic of Iran and Austrian Fund OeAD under IMPULSE Austria-Iran 15/2018 Program.

\bibliography{references}

\end{document}